\documentclass[aps, prl, twocolumn,showpacs,tightenline]{revtex4}

\usepackage{epsfig}
\begin{document}

\title{A Bell-type test of energy-time entangled qutrits}
\author{R.~T.~Thew$^1$}\email{Robert.Thew@physics.unige.ch}
\author{A.~Ac\'in$^{1,2}$}
\author{H.~Zbinden$^1$}
\author{N.~Gisin$^1$}
\affiliation{$^1$Group of Applied Physics, University of Geneva, 1211 Geneva 4, Switzerland}

\affiliation{$^2$Institut de Ci\`encies Fot\`oniques, Jordi Girona 29, 08034 Barcelona, Spain}

\date{\today}

\begin{abstract}
We have performed a Bell-type test for energy-time entangled qutrits. A method  of inferring the Bell violation in terms of an associated interference visibility  is derived. Using this scheme we obtained a Bell value of $2.784 \pm 0.023$, representing a violation of $34\,\sigma$ above the limit for local variables. The scheme has been developed for use at telecom wavelengths and using proven long distance quantum communication architecture to optimize the utility of this  high dimensional entanglement resource.
\end{abstract}


\maketitle


Bell Inequalities \cite{Bell64a} and other tests of non-locality \cite{Zeilinger99a} have a rich history in the evolution and understanding of quantum correlations and more specifically entanglement. This grew out of the EPR  position on the completeness of quantum mechanics \cite{Einstein35a}. However more recently we have been able to approach these tests from another perspective - that the violation of one of these Bell-type inequalities can be seen as a  witness of {\it useful}  entanglement \cite{Acin03a}. This has its roots in the qubit domain but has also been extended to qutrits, $d=3$  dimensional systems, in the context of  quantum complexity \cite{Bruckner02}. This is just one example of the current trend towards investigating higher-dimensional entanglement. This trend is motivated primarily by the promise of improved robustness of these states to noise and the possibility of  increased transmission rates for quantum key distribution schemes \cite{Bechmann00a,Cerf02a,Brub02a}. Perhaps more interestingly however is that with the increased complexity of these higher dimensional states comes the possibility of new quantum protocols that can capitalize on this useful entanglement.

There have been significant advances recently in the experimental investigation of higher dimensional entanglement. A range of schemes using various degrees of freedom have been put forth: a 4-photon polarization scheme generating states with spin-1 statistics \cite{Howell02a}; and a scheme incorporating lower order modes of orbital angular momentum (OAM) for photons producing qutrits \cite{Vaziri02a} have both performed Bell type tests; quantum state tomography has recently been performed for OAM entangled qutrits \cite{Langford03a} and interference experiments have shown time-bin entanglement up to $d=20$ \cite{Riedmatten03a}. 

Photonic entanglement is however best suited for quantum communication therefore if we are going to perform any protocol or distribute any entanglement over significant distance, we need to think about the architecture we use. A four-photon scheme, apart from obvious constraints due to polarization, is impractical as the encoding relies on all four photons being transmitted and detected. The OAM scheme will again have problems  with long distance fiber transmission predominantly due to dispersive effects between the different modes.  By contrast,  energy-time, and the similar time-bin, entanglement have a proven history over long distance \cite{Tittel99a,Thew02a,Marcikic03a} and the qutrit is encoded on a single photon. 

In this Letter we present the results for a Bell-type test, based on the inequality of Collins {\it et al.}  \cite{Collins02a} (CGLMP),  for energy-time entangled qutrits. The scheme is a natural extension of the  Franson arrangement for qubits \cite{Franson89a} and indeed the idea has previously been proposed \cite{Reck93a}. We also introduce a method and the associated constraints one needs to infer a Bell violation from an interference  visibility.  We discuss how these constraints correspond to the perception that higher dimensional entanglement is more robust and what this implies experimentally.


We will detail our approach to performing a Bell test momentarily, but first let's remind ourselves of the basic plot.  In theory the Bell test begins with the usual suspects, two
parties:  Alice and Bob, who are spatially separated. They share a
maximally entangled qutrit state and  they can choose between two different measurements of three outcomes. They determine various probabilities for the different  measurements and outcomes and calculate the relevant function to test the inequality.

\begin{figure*}[t]
\begin{center}
\epsfig{figure=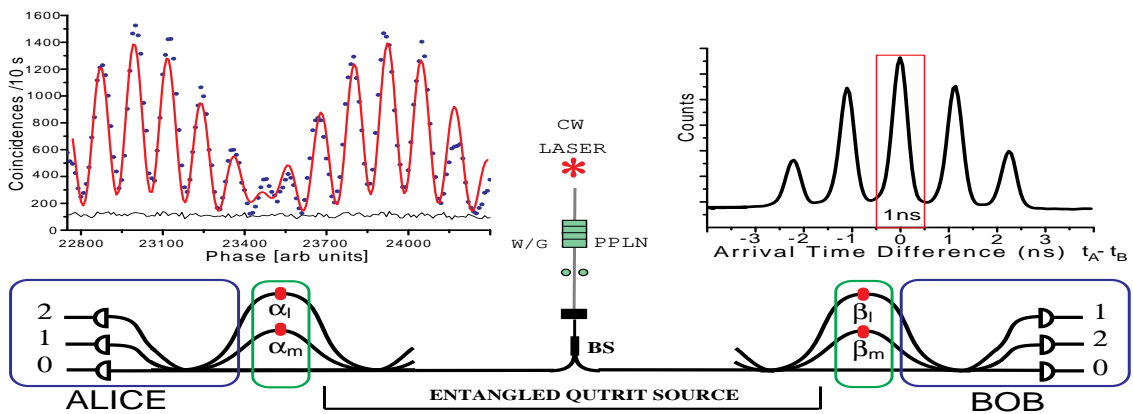,width=150mm,height=55mm}
\caption{Alice and Bob share entangled (energy-time at 1300\,nm) qutrits. Measurements are determined via variations in the path-lengths of their interferometers. There are 5 peaks in the arrival time histogram (shown on the right) due to different path combinations. Coincidence detection events in the central peak  project onto one of three orthogonal entangled qutrit states. These coincidences (shown on the left) vary as  a function of Alice and Bob's  phase vectors.}
\label{fig:schematic}
\end{center}
\end{figure*}

First we consider the experimental set-up used to perform this Bell test, see the schematic of Fig.\ref{fig:schematic}, to motivate a physical interpretation when we introduce the inequality. We use energy-time entangled photon pairs created at telecom wavelengths, via a PPLN waveguide \cite{Tanzilli01a}, and two three-arm interferometers \cite{Interferometers}  to generate and analyze entangled qutrits. For each interferometer we can define a phase vector consisting of the two independent phases, eg. the relative phases between the short-medium (m) and short-long (l), path-lengths. Coincidence measurements at the outputs of the interferometers project onto entangled qutrit states defined when the photons take the same path in each interferometer, short-short or medium-medium or long- long at Alice-Bob. 

For energy-time entanglement this is realized  by imagining that we have some detection time, say
 $t_0 + t_A $  for Alice, where $ t_A $ is the optical distance from the photon pair source to the detector, and a similar time  for Bob. Due to the long coherence length of the CW laser we do not have a well defined  $t_0$.  However with the hindsight of post-selection we can define a coherent superposition of three time-bin amplitudes, $t_0$, $t_0-\Delta \tau $, $t_0-2\Delta \tau $,  relative to the source that have well defined time differences. Thus we have our qutrit state prepared. Passing through the interferometer allows one to vary the two relative phases between these time-bins. As the  path-length differences  in the interferometer are $\Delta \tau $ and $2\Delta \tau$, exiting the interferometer corresponds to a fourier transform and a measurement in the transform basis defined by the post-selected $t_0$.  Bob does the same for his $t_0 + t_B$ and we post-select   entangled qutrits (The histogram central peak is centered at $t_B - t_A = 0$).

An arrival-time-difference histogram with five peaks, due to all the  possible  path combinations, like that inset on the right in Fig.\ref{fig:schematic}  is generated for each detector combination. Coincidence events in these central peaks correspond to projections onto states of the form:
\begin{eqnarray}
|\psi(j,k)\rangle &\propto & c_s|ss\rangle + c_me^{i (\alpha_m + \beta_m + \phi_{jk}^m)}|mm\rangle \nonumber \\  &&  \hspace{20mm}+  c_le^{i(\alpha_l + \beta_l + \phi_{jk}^l)}|ll
\rangle . \label{eq:qutrit}
\end{eqnarray}
Here $\alpha_m, \alpha_l$ and$\beta_m,  \beta_l$, represent the phases in Alice and Bob's medium and long interferometer arms. 
 $\phi_{jk}^m$ and $\phi_{jk}^l$ are multiples of $2\pi/3 $ which depend on the path taken by the photons in the interferometer and which output, $j,k \in \{0,1,2\}_{A,B}$, they take \cite{Zukowski97a,Thew03a}. 
 
 To have maximally entangled qutrits we need $|c_s|^2 = |c_m|^2 = |c_l|^2$. Experimentally, this relies on the symmetry of the fiber couplers. We require the splittting ratios to be 1/3:1/3:1/3, where an input signal at any one of the inputs is equally distributed in the three outputs. We use the same coupler for the interferometers input and output \cite{Interferometers}  and for both the interferometers the coupling ratios are within 5\% of this ideal value. We can then observe the  three orthogonal states corresponding to the three different coincidence detections, $0_A0_B$, $  0_A1_B$, $0_A2_B$ or  their cyclic permutations. When the phases are varied the coincidences vary as a function of Alice and Bob's phase vectors, a sample of which is inset on the left of Fig.\ref{fig:schematic} for a fixed but arbitrary ratio of medium and long phases. For more technical details concerning the experimental scheme we refer the reader to \cite{Thew03a}.

The CGLMP inequality is defined in terms of the measurement probabilities,, which we can define for  these states as $P^{{\rm Phase A, Phase B}}({\rm Result A, Result B})$. Due to the coupler symmetries we can satisfy the following constraints: $P^{mn}(0,0)= P^{mn}(1,1)= P^{mn}(2,2);  
P^{mn}(0,1)= P^{mn}(1,2)= P^{mn}(2,0);  P^{mn}(2,1)= P^{mn}(0,2)= P^{mn}(1,0)$. These relationships then simplify  the inequality such that,
 \begin{eqnarray}
I_3 &=& 3 [\{P^{11}(0,0) - P^{11}(0,1)\} + \{P^{21}(0,1) -
P^{21}(0,0)\}  \nonumber \\ && + \{P^{22}(0,0)  - P^{22}(0,1)\} +
\{P^{12}(0,0) -P^{12}(0,2)\}].\nonumber \\   && \le 2 \hspace{1cm}  ({\rm for \hspace{3mm} lhv})\label{eq:inequalitymedium}
\end{eqnarray}

For this inequality Alice and Bob have a choice of two phase settings each.  Each of these settings is a vector of two phases. We define phase vectors,  $A_i = (\alpha_{mi}, \alpha_{li})$,  for Alice's and $B_i = (\beta_{mi}, \beta_{li}$),  for Bob's.  The optimal {\it Bell} phase vectors are \cite{Durt01a, Collins02a}: 
 $ A_{1} = ( 0, 0); A_{2 }  = ( \pi/3, 2\pi/3); B_{1}= (\pi/6, \pi/3); B_{2} = ( -\pi/6, -\pi/3)$. The combination of the interferometers and these  phases realize a von Neumann measurement that is optimal in complete generality. 
Here we note that for each of these phase vectors we have the second phase equal to twice the first phase. We see in eq.(\ref{eq:qutrit}) that the state's phases depend on the sum of the phases in the two interferometers and one can also see that the vector sum of the Bell phases, $A_1+B_1, A_1+B_2...$ etc, retains this relationship which is an important constraint that we will come back to momentarily.  With these settings the coincidence probabilites are further constrained such that we have each of the four bracketed terms in eq.(\ref{eq:inequalitymedium})  equal.  The inequality thus reduces to
\begin{eqnarray}
I_3 &=& 12[P^{11}(0,0) - P^{11}(0,1)]  = 12[\frac{4 + 2\sqrt{3}}{27}-\frac{1}{27}] \nonumber \\
& \approx & 2.872... \label{eq:inequalityshort}
\end{eqnarray}

We now consider imperfections in the system.  The simplest way to do this is to introduce noise to the system and when looking at Bell inequalities one normally uses a symmetric noise model. One can then characterize a measure on the system in terms of its robustness to the admixture of noise.  Consider the state
  \begin{eqnarray}\label{eq:werner}
      \rho = \lambda|\psi\rangle \langle \psi | + (1-\lambda)\frac{{\cal I}_{9}}{9}
  \end{eqnarray}
  where $|\psi\rangle$ is a maximally entangled pure qutrit state and
  ${\cal I}_{9}$ is the Identity operator for the entangled qutrit space, $d=9$. The CGLMP inequality scales simply with this mixing parameter, $ \lambda$, such that  $ I(\rho) = \lambda I_{3}$ \cite{Collins02a}. Inverting this gives the critical mixing value such that the inequality is violated. This inequality is defined for arbitrary finite dimensions and one finds for $d=2$, $\lambda_2^c = 1/\sqrt{2} \approx  0.707$. For qutrits, $d=3$,  the critical mixing value is lower,  $\lambda_3^c = (6\sqrt{3}-9)/2 \approx 0.696...$, and  hence it is more robust with respect to noise. The CGLMP inequality reveals the robust nature of higher dimensional entanglement as the amount of noise that can be added to the system increases with the dimensions of the system \cite{Collins02a}.  

If we assume a symmetric noise model as in eq.(\ref{eq:werner}) for our experiment then we can determine the coincidence probabilities as a function of the two phases, in the medium and long arms of the two interferometers, and the mixing parameter, $\lambda$,
\begin{eqnarray}
   P_{jk}& \propto & 3 + 2\lambda [\cos(\alpha_m + \beta_m +\phi_{jk}^m) + \cos(\alpha_l + \beta_l + \phi_{jk}^l)\nonumber \\  &&\hspace{5mm} + \cos(\alpha_m + \beta_m - \alpha_l - \beta_l + \phi_{jk}^m-\phi_{jk}^l)]. \label{eq:expprob}
\end{eqnarray}
In practice we do not take measurements at fixed phase settings, instead we continuously scan the phases. This is done in a controlled manner such that we always have the long phase twice that of the medium, as is the case for the Bell phase vectors, as previously mentioned. Hence the coincidence probability becomes a function of just the one phase and the mixing parameter as in the qubit case. This is confirmed experimentally by looking at the interference events associated with the satellite peaks in the arrival-time histogram. It can be shown that if the phase is varying at the same rate in both of these peaks then we have the desired factor of two relating the two phases \cite{Thew03a}. This means that we have the symmetry simplified to the level of eq.(\ref{eq:inequalityshort}). It also means that we can use a fitting function based on  eq.(\ref{eq:expprob}) to directly determine $\lambda$  and hence also determine the value for the CGLMP inequality.

We have used interferometric methods to generate entangled qutrits and as such we would like to analyze the system using standard interferometric techniques, like interference  visibilities. This approach is well understood and often used when characterizing qubit schemes. In the case of qubits the mixing parameter, $\lambda$, corresponds directly to the visibility and hence a visibility greater than $\lambda_2^c$ implies the state is capable of violating the inequality.  Of course, to infer this violation one must be able to satisfy various constraints that depend on the symmetry of the system.

In the case of qutrits, and states of higher dimensions in general, we then can define the visibility  in the following manner. Consider eq.(\ref{eq:werner}), with the Identity and the entangled state temporarily defined on a $d$-dimensional space, such that we have the following result,
 \begin{eqnarray}
V(d) = \frac{R_{max}-R_{min}}{R_{max}+R_{min}} = \frac{d \lambda}{2 + \lambda(d-2)} \label{eq:vis}.
 \end{eqnarray}
Here one can imagine that for some combination of detectors we have perfect correlations and expect a maximum coincidence rate of  $R_{max}=\lambda + (1-\lambda)/d$, the first term due to the $d$ possible outcomes and the second term due to the noise. If we are perfectly uncorrelated we expect  $R_{min}=(1-\lambda)/d$ due only to the noise.  
\begin{figure}
\begin{center}
\epsfig{figure=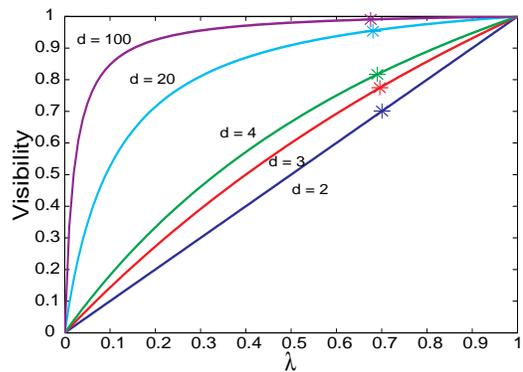,width=70mm,height=50mm}
\caption{The relationship between the mixing parameter $\lambda$ and the visibility for various dimensions. Also shown are the critical values for $\lambda$ (*)  to violate the Bell inequality. }
\label{fig:VvsDim}
\end{center}
\end{figure}
To satisfy these conditions in our experiment we need the same symmetry constraints on the interferometer couplers and the 2:1 phase constraint, as one needed for the Bell-test.

In Fig. \ref{fig:VvsDim} we show this function and clearly see that while the  critical values for $\lambda$, denoted by (*),  to violate the Bell inequality do decrease, marginally, with the dimensions, the visibility becomes significantly more robust. For the case of qutrits we have $V(3) = 3\lambda /(2+\lambda)$ which implies a visibility greater than 0.775..., even though we have added more noise than in the qubit case.  Experimentally, this can be reconciled, in part, as a result of  the standard Bell model and symmetric noise which adds the noise to irrelevant degrees of freedom.

\begin{figure}
\begin{center}
\epsfig{figure=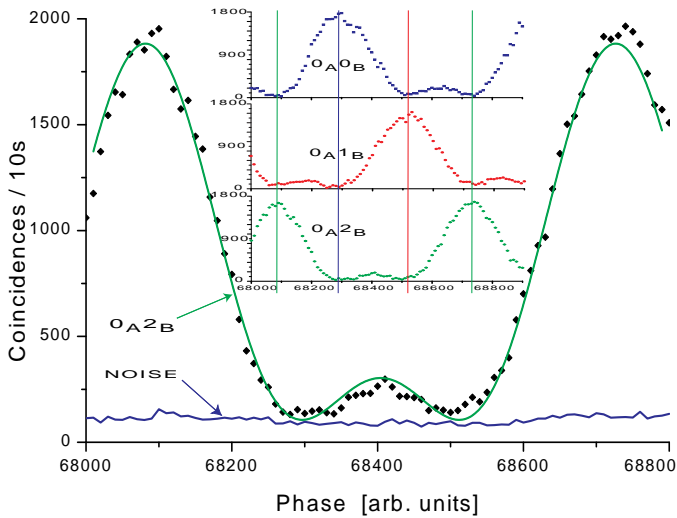,width=90mm,height=70mm}
\caption{ Raw coincidence count rates, circles, as the phases are varied under the Bell phase constraint. Also shown is a fit using eq.(\ref{eq:expprob}) and the noise level.  Inset are the  three (net) orthogonal results.}
\label{fig:fringe}
\end{center}
\end{figure}

In Fig.\ref{fig:fringe} we see the coincidence counts as a function of one phase, with the other in this fixed 2:1 relationship.  A least-squares fit of this data returns a value $\lambda = 0.848 \pm 0.008$. From this we can then directly calculate the inequality, $I_{exp} = \lambda I_{3} = 2.436 \pm  0.023$ which corresponds to a violation of the inequality by $19\,\sigma$. If on the other hand we wish to directly interpret this in terms of the visibility we can use eq.(\ref{eq:vis}) and obtain, $V(3) = 0.893 \pm 0.006$. The critical visibility value for qutrits is  0.775... and we find from the visibility a violation of $19 \sigma$ as we would expect from the previous result. This is the raw result which includes the background noise counts as well as those due to the correlated photons. We can directly, and concurrently, measure this noise, also shown in Fig.\ref{fig:fringe},  by looking at detection  events that arrive outside of these five peaks in the histogram of Fig.\ref{fig:schematic}.  If we subtract this noise and look at the net results we find  $\lambda_{net} = 0.969 \pm 0.008$. Which in turn gives us a net Bell value of $I_{exp-net} = \lambda_{net} I_{3} = 2.784 \pm  0.023$  and a net visibility of $V(3)_{net} = 0.979 \pm 0.006$ with a violation of the inequality by $34\,\sigma$.

In the inset of Fig.\ref{fig:fringe} we have shown the net coincidence counts for the three orthogonal outputs, corresponding to  coincidence detections at $0_A0_B, 0_A1_B, 0_A2_B$, the raw results for the  curve $0_A2_B$ are shown in the main figure. We clearly see the signature three-way symmetry for the entangled qutrits with the maxima evenly separated in phase space. In terms of the correlations we also see that for each maxima for one output we have minima, almost at the noise level, at the other two as we would expect given the high visibility.

In this Letter we have presented the results for a Bell-type test for energy-time entangled qutrits achieving violations of 19 and 34$\,\sigma$ for the raw and net results, respectively. We have also derived a simple means of determining this violation in terms of interference  visibility. We have approached both the experimental design and the Bell test itself with the generation and characterization of a source of useful qutrit entanglement in mind.   The high signal to noise level for the raw results reinforces the utility of this arrangement and its suitability to high dimensional  long distance quantum communication.

The authors would like to thank the University of Nice for supplying the PPLN waveguide and gratefully acknowledge financial support from the Swiss NCCR "Quantum Photonics" and the EU IST-FET project RamboQ.


\begin{references}
\bibitem{Interferometers} We have shown Mach-Zender interferometers which are intuitively easier to understand, however, in reality, we use  Michelson interferometers which are more practical.
\end{references}
\end{document}